# Optimizing site-specific specimen preparation for Atom Probe Tomography by using hydrogen for visualizing radiation-induced damage

*Aparna Saksena*[*,1], *Binhan Sun*[1,2], *Xizhen Dong*[1], *Heena Khanchandani*[1], *Dirk Ponge*[1], *Baptiste Gault*[1,3]

[1] Max-Planck-Institut für Eisenforschung GmbH, Max-Planck-Straβe 1, Düsseldorf 40237, Germany

[2] Key Laboratory of Pressure Systems and Safety, Ministry of Education, School of Mechanical and Power Engineering, East China University of Science and Technology, Shanghai 200237, China

[3] Department of Materials, Royal School of Mines, Imperial College London, Prince Consort Road, London, SW7 2BP, UK

*a.saksena@mpie.de

## Abstract

Atom probe tomography (APT) is extensively used to measure the local chemistry of materials. Site-specific preparation via a focused ion beam (FIB) is routinely implemented to fabricate needle-shaped specimens with an end radius in the range of 50 nm. This preparation route is sometimes supplemented by transmission Kikuchi diffraction (TKD) to facilitate the positioning of a region of interest sufficiently close to the apex. Irradiating the specimen with energetic electrons and ions can lead to the generation of vacancies and even amorphization of the specimen. These extrinsically created vacancies become crucial for probing the hydrogen or deuterium distribution since they act as a strong trap. Here, we investigated the feasibility of site-specific preparation of a two-phase medium-Mn steel containing austenite (fcc) and ferrite (bcc). Following gaseous charging of APT specimens in deuterium ($D_2$), clusters enriched by up to 35 at.% D, are imaged after Pt deposition, conventional Ga-FIB preparation, and TKD conducted separately. These D-rich clusters are assumed to arise from the agglomeration of vacancies acting as strong traps. By systematically eliminating these preparation-induced damages, we finally introduce a workflow allowing for studying intrinsic traps for H/D inherent to the material.

**Introduction**

Hydrogen is viewed as a carbon-free, clean alternative to replace the currently used non-renewable sources such as coal, petroleum, and natural gases [1]. It is the lightest element, is ubiquitous and has a high diffusivity [2]. Therefore, its interaction with materials cannot be avoided. Only a few ppm of H can lead to an abrupt loss in ductility and strength causing premature failure of the material [2-5]. This phenomenon is known as hydrogen embrittlement. Although first discovered in the 19$^{th}$ century, to date, the mechanism remains not fully understood, with one primary challenge of accurately probing H distribution inside materials.

Atom probe tomography (APT) [6] has garnered a lot of interest because of its capability to reveal the spatial distribution of the elements with a sub-nm resolution; especially regarding the direct observation of H, APT has been considered a vital tool [7-9]. However quantitative measurement of H via atom probe requires extra measures to differentiate between the residual H originating from the analysis chamber [10-12], and from the H initially inside the sample. Therefore, because of its low isotopic abundance, deuterium (D) is sometimes used to charge the specimens [7, 9, 13], which can facilitate but not guarantee quantitative analysis [14]. Furthermore, due to the high diffusivity of H/D in most materials, following charging, the specimens may require fast quenching to cryogenic temperature to kinetically hinder the outward diffusion of H/D [9, 15].

A few studies have reported the spatial distribution of D and its enrichment in carbides, precipitates, dislocations, and grain boundaries via APT [7, 9, 16-21]. For all these, specimens were prepared using electrochemical polishing, which typically does not allow for the analysis of a pre-selected microstructural feature of interest. Breen *et al.* also reported the incorporation of H in the specimen due to electropolishing by up to 80%, which may already saturate all of the sites with H even before deuteration [13].

For studying specific microstructural features of interest in the limited volume analysed by APT, specimen preparation via a focused ion beam (FIB) is well established [22, 23]. This technique is most efficient to ascertain that a microstructural feature is present in the small information volume that is analysed by APT. This includes dislocations [24], stacking faults [25], and grain boundaries [26-28], which are extracted from the bulk of the sample by using a FIB and mounted on a support. Frequently, the specimen preparation is supplemented with transmission Kikuchi diffraction (TKD) [23, 27]. Such a technique becomes especially crucial to probe the interaction of lattice defects with H [29], when they are heterogeneously and sparsely distributed in the material.

For studying hydrogen, electrochemical charging of APT specimens prepared by FIB was reported to be challenging [30], and gas charging was proposed as an alternative [31]. FIB, however, is known to introduce structural damage to the specimen [22, 32]. How H interacts with structural damage introduced by the interaction of the energetic ion beam with the sample has not yet been studied in detail. While Ga or Xe implantation from the FIB is typically visible in APT, structural damage in the form of isolated or clustered vacancies is not and *ab initio* calculations suggest that vacancies are a strong trap for H in Al [33, 34], Pd [35], bcc Fe [36] and fcc Fe [37]. Electron imaging in the transmission electron microscope (TEM) [38] or transmission Kikuchi diffraction (TKD) [39], sometimes used to characterize microstructural features in APT specimens [26], can induce structural damage in the form of isolated or clustered vacancies. Gault *et.al.,* [40] recently showed how hydrogen segregation following gas charging of APT specimens revealed this damage. This work emphasized how critical it was to minimize radiation-induced damage for probing the distribution of H/D in the materials, as any extrinsically induced trapping sites would hinder the investigation of a material's intrinsic traps [22].

Here, we investigate the feasibility of site-specific specimen preparation for probing H/D by APT in a two-phase medium-Mn steel consisting of ferrite and austenite. This alloy exhibits excellent strength and ductility combination but is particularly prone to hydrogen embrittlement [4, 41, 42]. We systematically study the influence of the main steps and parameters used during specimen preparation, namely Pt deposition, the use of Ga-FIB or Xe-FIB, and the use of TKD. Following gaseous charging of APT specimens, deuterium-rich clusters are used as a tracer for the damage. By systematically optimizing preparation steps, we finally introduce a workflow that allows us to minimize damage from both the electron and ion beams, and showcase imaging of the distribution of charged deuterium at structural defects.

**Experimental Methods**

We selected a medium Mn steel with a chemical composition 0.2C-10.2Mn-2.8Al-1Si (in wt.%). This material is intercritically annealed to achieve an equal volume fraction of ferrite (α) and austenite (γ). It has been reported to undergo a dramatic loss in ductility in the presence of H and therefore it is vital to reveal the H traps in such a material system. More details can be found in [43].

The APT specimens were prepared using the procedure described in [22], using a FEI Helios NanoLab 600i dual beam FIB or a FEI Helios dual beam Xe-plasma FIB. The support is a 316 stainless-steel grid described in [31], held in a home-made holder described in [38]. APT specimens were transported in an ultra-high vacuum suitcase [44] to avoid exposure of the specimen to the atmosphere between the gas charging chamber called the Reacthub module (RHM) [31] and the atom probe.

For deuterium charging, specimens were placed in a 25 kPa $D_2$ atmosphere. The stainless-steel grid allows focusing an infrared laser on the specimens to apply a temperature in the range of 250 °C – 300 °C. This temperature is selected to avoid any microstructural changes in the specimen, while simultaneously reducing the kinetic barrier for deuterium diffusion into the specimen. After 6 hours, the infrared laser is stopped and the chamber is evacuated to $5\times10^{-6}$ mbar, to remove the introduced $D_2$. The specimens were transferred to the suitcase where they were cryogenically cooled. The duration between the stopping of the infrared laser and transferring the specimen to the suitcase is approximately 5 minutes, while it is expected that it takes further 15 minutes for the specimens to reach 80 K after being placed in the cryogenically cooled suitcase, based on the similarity in the cryo-stage used for the commercial atom probes. This strategy likely enables us to probe only deeply trapped D. Subsequently, specimens are transported in the suitcase, to a local electrode atom probe (LEAP 5000 XS) for analysis. In the LEAP, the specimens were field evaporated in voltage pulsing mode by collecting 5 ions in every 1000 pulses at 60 – 70 K base temperature. A pulse repetition rate of 200 – 250 kHz and a pulse fraction of 15% was applied. The collected data was reconstructed using AP Suite 6.1.

To probe the feasibility of a site-specific preparation for investigating H/D via APT, five workflows were examined as shown in Figure 1. Workflow 1 involved specimen preparation by using a Pt protective layer deposited using an ion beam, followed by standard lift out procedure using a Ga ion FIB. TKD was employed to position the region of interest, closer to the specimen apex. TKD was eliminated for Workflow 2. In Workflow 3, additionally, the Pt protective layer deposited via the ion beam was replaced by the Pt protective layer deposited via the electron beam. In Workflow 4 we exposed the specimen, prepared using workflow 3, by 30 kV electron beam, similar to that employed for TKD, for 3 minutes. In Workflow 5, the specimens were prepared by using a Pt protective layer deposited using the electron beam. The Ga ion beam was replaced by the Xe ion plasma for the standard lift out. No TKD was employed for this workflow. All specimens are cleaned using a low kV ion beam before APT and gas charging.

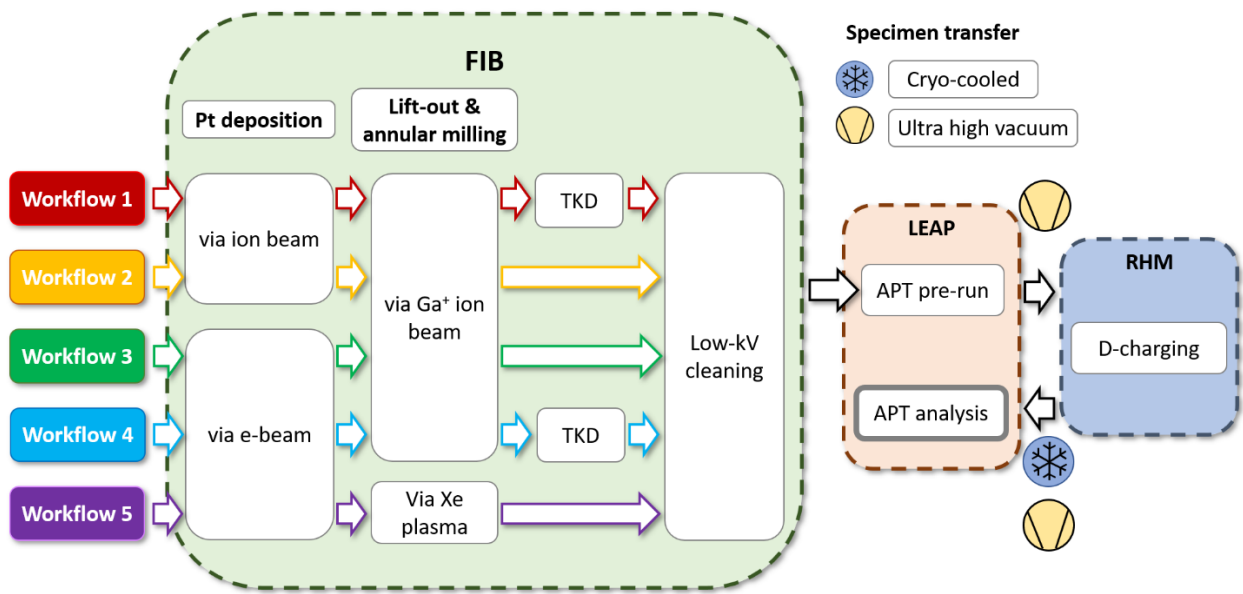

*Figure 1: Specimen preparation and analysis workflows explored here to enable analysis of H/D by APT.*

## Results and Discussion

### Workflow 1

The differences in mobility and solubility of H between α and γ make the α–γ phase boundary a region of interest to understand the material's response to H exposure. We targeted this interface for site-specific specimen preparation. To protect the region of interest from ion beam irradiation, a Pt protective layer was deposited using an ion beam energy of 30 kV prior to lifting out. The α–γ phase boundary was brought closer to the apex of the specimen with the help of an iterative process of TKD-guided annular milling. TKD patterns were obtained by using an electron beam accelerated to 30 kV. The phase maps of the specimens are shown in Figure S1 (supplementary information). After ensuring the presence of the α–γ phase boundary, the specimens were cleaned by using a low energy (5 kV) $Ga^+$ ion shower.

Before deuteration, the specimens were cleaned by collecting 2 to 3 million ions in the atom probe. The obtained mass spectrum is shown in the supplementary information where the peak at 1 Da belongs to $H^+$. These pre-deuteration measurements were performed for all the subsequent measurements to compare the field conditions during the APT measurements, before and after deuteration.

After deuteration, APT was performed to quantify the distribution of D in the specimen. Figure 2 a) shows the reconstruction of a representative measurement where the distribution of Fe and Mn can be visualized in α and γ. A 6.5 at.% Mn iso-surface is used to delineate the α–γ phase boundary. The corresponding mass spectrum (focusing on 1 to 3 Da), is shown in Figure 2 b) where peaks at 1, 2, and 3 Da can be observed. The peak at 1 Da corresponds to $H^+$, however, the peaks at 2 and 3 Da may respectively correspond to $D^+$ and $DH^+$ or $H_2^+$ and $H_3^+$.

Since the relative amplitude of H-containing molecular ions depends on the electric field during the measurement [45], further analysis is required to clarify the origin of these peaks. From the post-ionization model [46], it is well-accepted that the charged state ratio of an element is representative of the field condition during the measurement. Here we compare the relative abundance of $Al^{+3}$ with respect to $Al^{+2}$, appearing at 9 and 13.5 Da respectively. The electric field is estimated to be between 36.2 V/nm – 37.4 V/nm, with considerable overlap between the pre-deuterated and deuterated specimens (supplementary information). With similar field conditions, the pre-deuterated sample showed only a peak at 1 Da whereas the deuterated specimen showed peaks additionally at 2 and 3 Da. Therefore, these peaks are considered to be $D^+$ and $DH^+$ respectively.

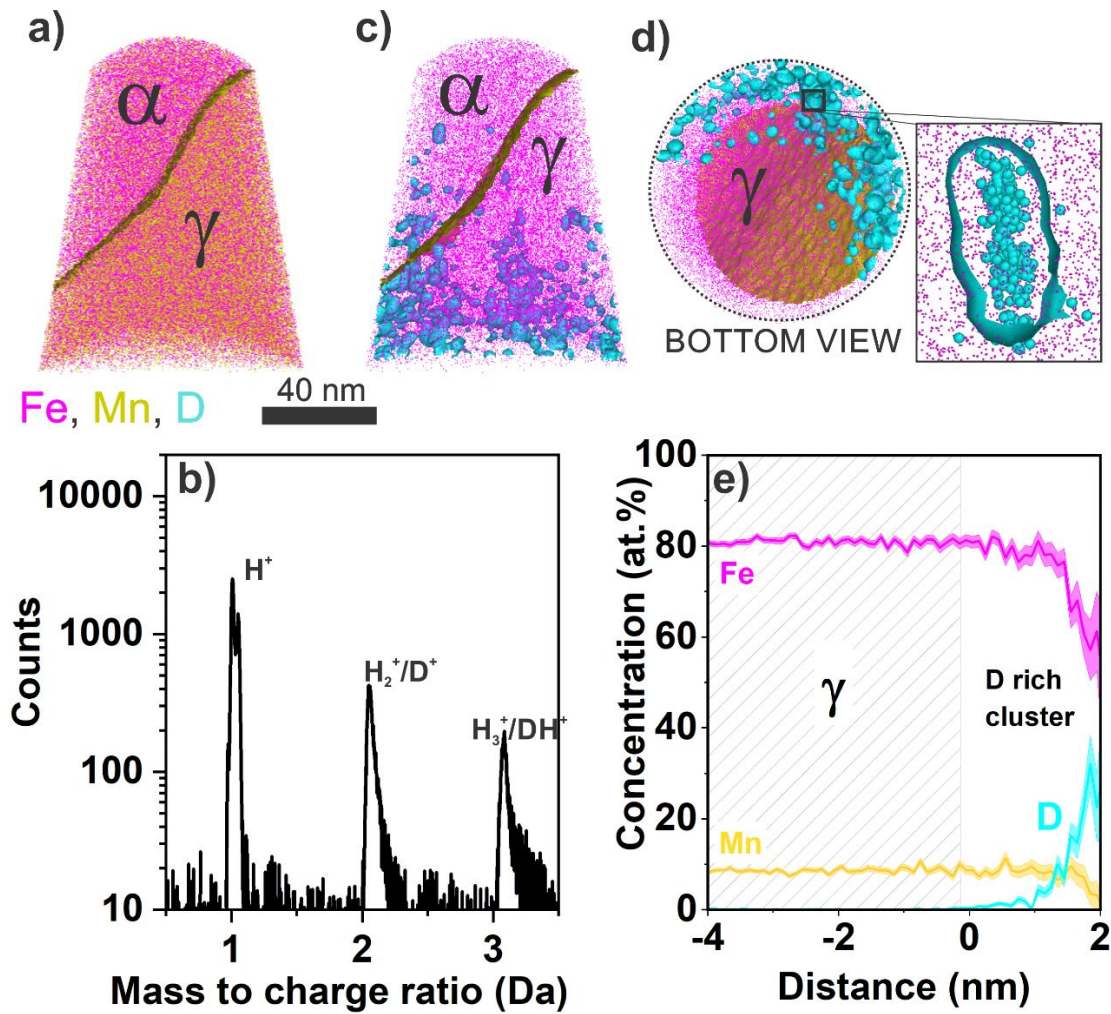

Figure 2: a) APT reconstruction of the specimen prepared by conventionally employed strategy for site-specific lift out using a Ga+ ion FIB and further characterized by TKD during annular milling. The distribution of Fe and Mn is shown where a 6.5 at.% Mn iso-surface shows the phase boundary. b) shows the mass spectrum obtained from the measurement of the deuterated specimen. c) and d) 1 at.% D iso-surface, in cyan, shows the distribution of D in the reconstructed volume where d) shows the top view of the reconstruction. e) shows the proximity histogram of the interface between a D-rich region and γ.

A set of 1 at% D iso-surfaces, in blue, superimposed to the point cloud in Figure 2 c) evidences D-enriched regions that are similar to the recent report by [40] of H/D filled voids. The top view of the reconstruction in Figure 2 d) suggests that these voids are in higher density towards the top-right side of the specimen. A composition profile for a representative filled void, calculated in the form of a proximity histogram [47], and plotted in Figure 2 e) reveals an enrichment of up to 35 at.% D. These voids likely originate from the clustering of vacancies from sub-surface cascades resulting from ion or, indirectly, electron irradiation. Vacancies are strong traps for H/D [48] and therefore, voids show strong enrichment of D [37].

It should be noted that these D-enriched voids are present in both phases. However, many factors directly influence the extent of damage for each phase such as the shape of the specimen, the position of the phase with respect to each other and with respect to the electron and ion beam, the volume of the specimen that is irradiated and the measured volume that is limited by the field-of-view of the LEAP. Therefore, a direct and quantitative assessment of the damage for the different phases is not directly possible.

## Workflow 2

To minimize these irradiation-induced damages, as a first step, we eliminated TKD during specimen preparation. Since, the material, due to its ultrafine microstructure, contains a high density of phase boundaries, there is still a high probability to capture such a feature in our measurements. The specimen was prepared by first depositing a Pt protection layer via a 30 kV-$Ga^+$ ion beam followed by the same lift-out procedure and annular milling using a 30 kV-$Ga^+$ ion beam, and a final low-kV cleaning step using 5 kV-$Ga^+$. Figure 3 a) is the APT reconstruction from the analysis after deuteration. A 6.5 at.% Mn iso-surface highlights the α–γ phase boundary. The corresponding mass spectrum, in Figure 3 b), shows D$^+$ and DH$^+$ at 2 and 3 Da respectively. Similar D-enriched voids are highlighted in Figure 3 c) by using a set of 0.7 at.% D iso-surfaces. These voids are enriched by up to 20 at% D.

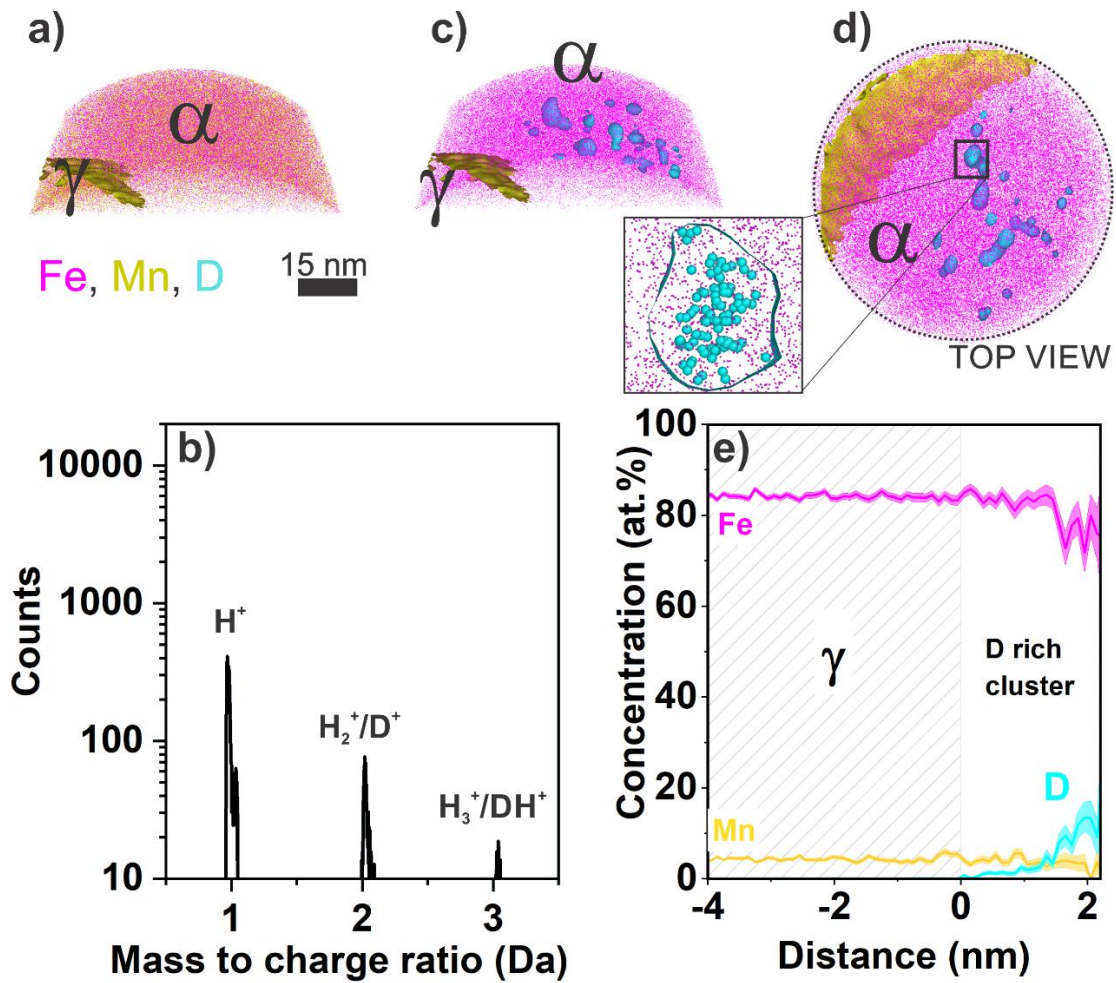

*Figure 3: a) shows the APT reconstruction of the specimen prepared using the conventional site-specific lift-out where TKD was eliminated. The distribution of Fe and Mn is shown where a 6.5 at.% Mn iso-surface is used to show the phase boundary. b) mass spectrum obtained during the APT measurement. c) and d) shows the distribution*

*of D by using a 0.7 at.% D iso-surface. e) shows the proximity histogram of the interface between the highlighted D-rich region (inset of d) and ferrite.*

## Workflow 3

Although this specimen preparation approach is thought to cause minimum damage, we still observe strongly trapped D, clustered, in the APT reconstruction in Figure 3, indicating that preparation-induced damage is still prevalent. Pt deposition by using the Ga-ion beam has been reported to induce damage over 60 nm sub-surface, whereas damage from Pt deposition by using the electron beam was below 10 nm in an Al alloy [49]. Therefore, the preparation technique is further optimized by depositing Pt to protect the region of interest by using a 5 kV-electron beam instead of the previously used 30 kV ion beam. A similar lift-out procedure is performed, followed by annular milling and cleaning, using the Ga$^+$ ion beam.

Following deuteration, specimens were analysed by APT, and the corresponding reconstructed data is displayed in Figure 4 a), with the α–γ phase boundary marked by a 6.5 at.% Mn iso-surface. The corresponding mass spectrum shows a peak at 2 Da indicating D incorporation in the specimen after gas charging. Figure 4 b) shows the distribution of D in the analysed volume by using 0.5 at.% D iso-surface, which highlights D-enriched voids at the periphery of the specimen as visible in Figure 4 c). Though we still observe the preparation-induced damage, it is considerably reduced compared to workflows 1 and 2.

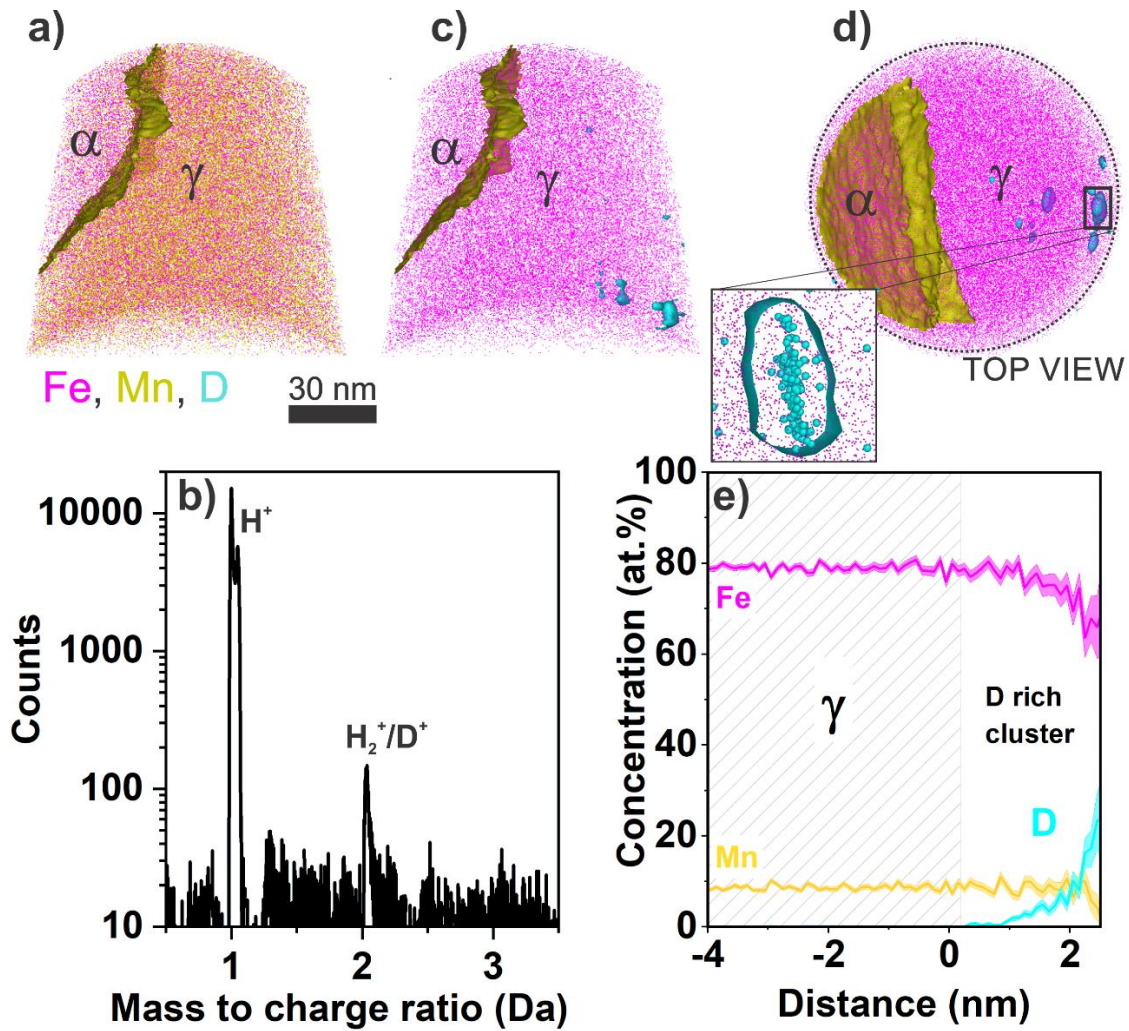

*Figure 4 a) shows the APT reconstruction of the specimen prepared using the Ga+ ion beam. A 6.5 at.% Mn iso-surface shows the phase boundary between ferrite and austenite. b) shows the mass spectrum obtained during the APT reconstruction. c) and d) shows the distribution of D by using a 0.5 at.% D iso-surface. e) shows the proximity histogram of the interface between the D-rich region and γ.*

Workflow 4

The D-enriched voids in Workflow 1 resulted from damage caused by the TKD and ion-beam-based Pt-deposition. To assess the feasibility of TKD-guided specimen preparation, specimens were prepared following e-beam Pt-deposition. The annular milling was performed by a 30 kV-Ga$^+$ ion beam. The specimens were exposed to a 30 kV-electron beam in a configuration similar to TKD, i.e. where the specimen is inclined at 38° to the electron beam. The beam was rastered over the specimens for 3 minutes. Specimens were then cleaned using a 5 kV-Ga ion beam.

After deuteration, APT was performed and the corresponding reconstruction is shown in Figure 5 a). The distribution of Fe and Mn is shown and a 5 at.% Mn iso-surface is used to demarcate the α–γ phase boundary. The mass spectrum is plotted in Figure 5 b), showing peaks at 2 and 3 Da corresponding to D$^+$ and DH$^+$. The distribution of D in Figure 5 c) shows a high density of D-enriched voids. Figure 5 d) shows the top view of the reconstruction, further highlighting the distribution of these voids towards one side of the specimen.

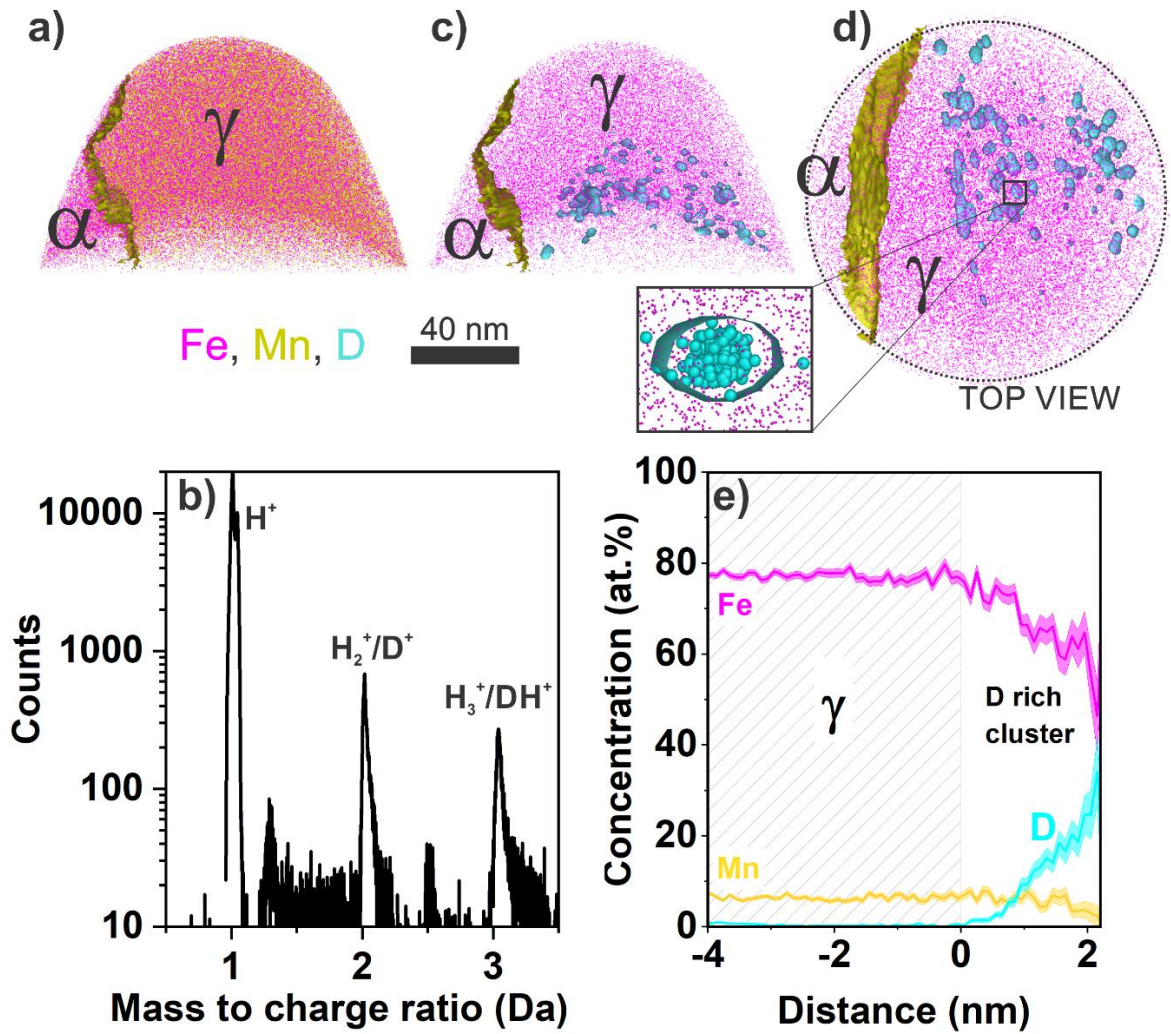

*Figure 5: a) APT reconstruction of the specimen exposed to an electron beam of 30 kV energy for 3 minutes. The distribution of Fe and Mn is shown in α and γ. A 5 at.% Mn iso-surface is used to demarcate the α–γ phase boundary. b) shows the mass spectrum obtained from the APT measurement. c) and d) shows the distribution of D by using a 1 at.% D iso-surface. e) shows the proximity histogram between the D-rich region and γ.*

Workflows 3 and 4 reveal the relative impact of TKD on the formation of D-enriched voids. This can be explained by the angle of incidence between the electron beam and the specimen [50]. For annular milling, the Ga$^+$ ion beam is mostly parallel to the specimen shank. Most Ga$^+$-induced damage is observed when the specimen gets thicker further down the shank, i.e. where is it no longer nearly parallel, but slightly inclined with respect to the ion beam. Whereas the electron beam in TKD geometry is at 38° inclination to the specimen, creating a larger interaction volume [50]. Although the momentum transfer due to a Ga ion and an electron of the same energy is considerably different, because of the mass of the incident species, the recoil of lightweight surface species, such as H and C in the FIB chamber, was proposed to cause the structural damage [40]. From our observations, even 3 minutes of exposure to a 30 kV-electron beam, incident at 38° to the specimen, is sufficient to create considerable damage in the specimen. This makes it difficult to employ a TKD-guided specimen preparation for investigating H/D distribution, intrinsic to the specimen.

## Workflow 5

The damage induced by 30 keV Ga$^+$ and Xe$^+$ ions implanted in the same material was reported to differ [51, 52]. Therefore, to further reduce damage from annular milling, a Xe plasma FIB was employed for the lift-out, annular milling, and low-kV cleaning at 5 kV. After deuteration, APT was performed. Figure 6 a) is the corresponding reconstruction, in which only austenite is captured, and Figure 6 b) plots the mass spectrum. No measurable amount of D is detected, indicating the absence of any strong traps in the analysed volume: we have now devised a workflow that minimizes preparation-induced damage.

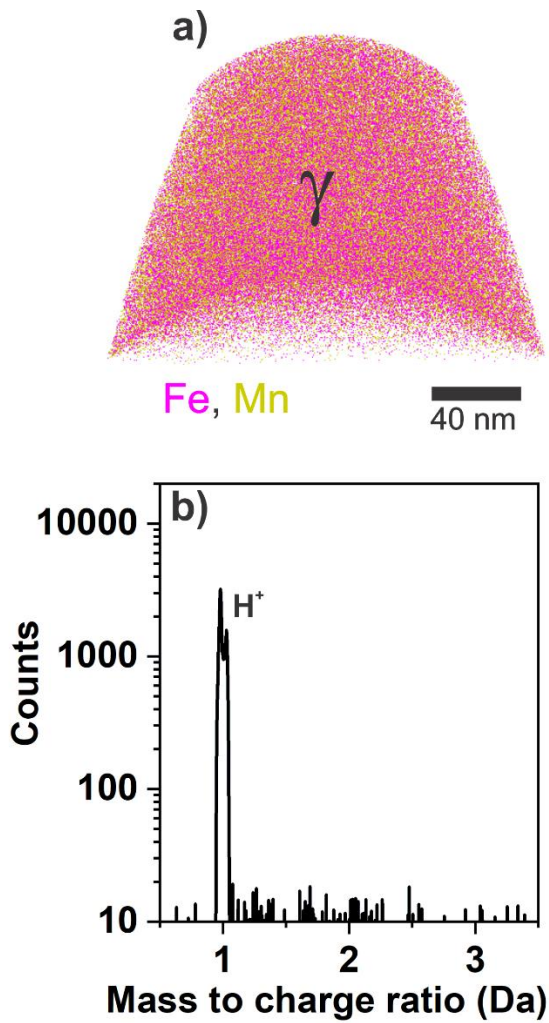

*Figure 6: shows the APT reconstruction of the specimen prepared using a Xe plasma FIB. b) shows the mass spectrum obtained during the measurement, showing no incorporation of D.*

Summary and discussion

By using a systematic elimination approach, summarized in Figure 1, we managed to optimize the specimen preparation workflow, leading to a minimization of the damage to finally prepare specimens that did not contain any extrinsically- and artificially-introduced trapping sites. We have demonstrated that a site-specific specimen preparation route, although possible, requires much more vigilance to probe the intrinsic behaviour of a material under H/D. This involves an electron beam to avoid the primary damage from the ion beam during Pt-deposition of the protective cap prior to lift-out, using a Xe-beam if possible, and avoiding TKD.

It must be noted that the damage associated with the ion and electron radiation should not be specific to the currently probed material system but is crucial for any material that is studied to analyse the microstructural distribution of H/D. It is fairly clear that the vacancy clusters strongly interact with H/D. Therefore, it could be a reliable strategy to track such defects in a volume, by chemically decorating them with H/D. This could be specially interesting for better understanding vacancy assisted mechanisms.

Using this optimized workflow (#5), and after deuteration, a final APT measurement was performed, and Figure 7 a) shows the distribution of Fe and Mn is shown, along with a 6.5 at.% Mn iso-surface to highlight the α–γ phase boundary. The corresponding mass spectrum, in Figure 7 b), shows a peak at 2 Da corresponding to $D^+$, indicating the incorporation of D into the specimen. As expected, no D-enriched voids are observed. A linear feature in the ferrite, adjacent to the phase boundary, is selectively highlighted by a 0.5 at.% C iso-surface as readily visible in Figure 7 c). This linear feature is ascribed to a dislocation enriched with C. Figure 7 d) plots a proximity histogram calculated as a function of the distance to this iso-surface, only the C and D concentration profiles is displayed. We see enrichment of D of up to 0.3 at.%, at the dislocation core, in conjunction with the C enrichment. Dislocations have been shown to trap D in the literature which is in good agreement with our observations [17, 18, 48].

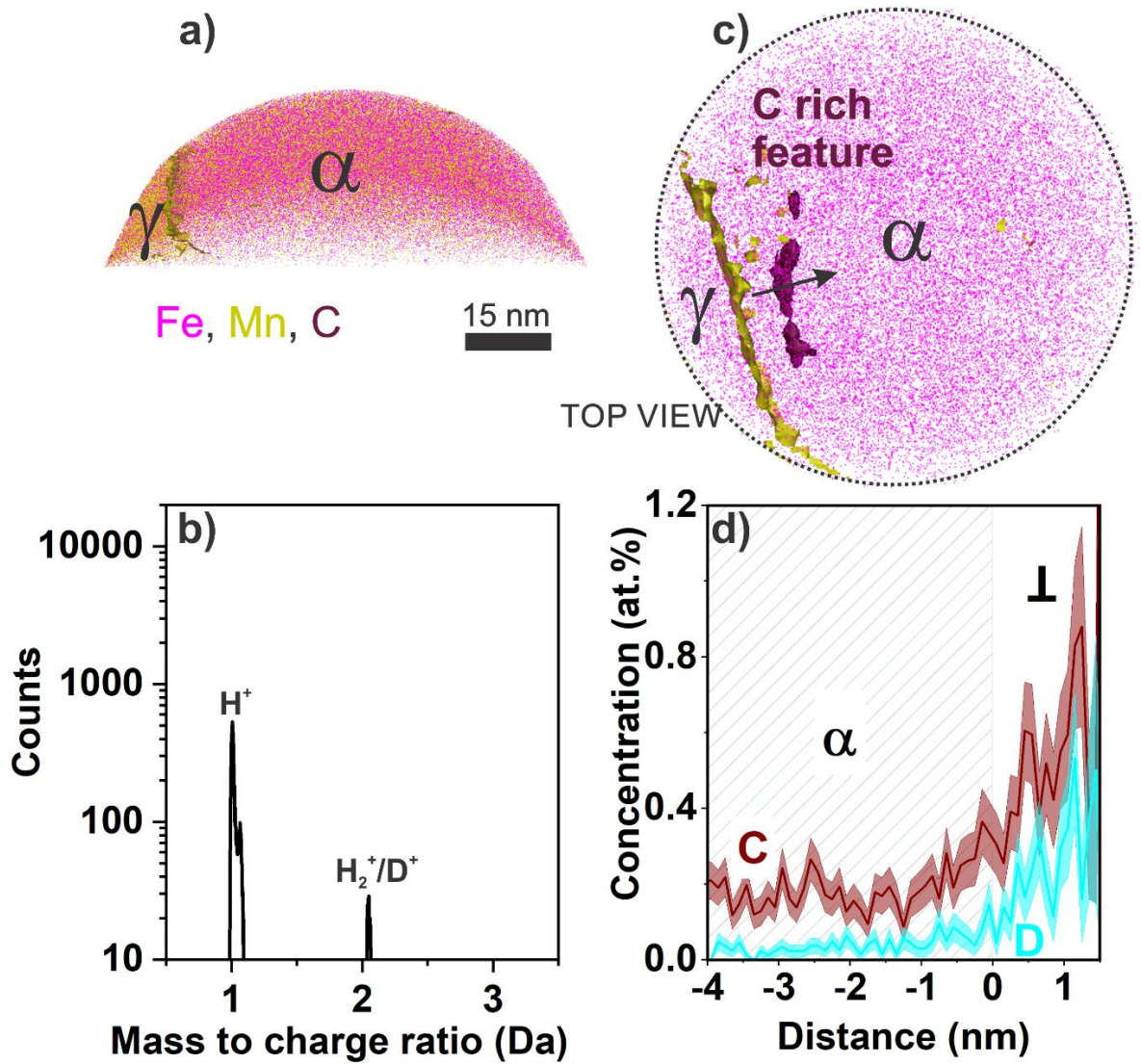

*Figure 7: a) The APT reconstruction of the specimen prepared using a Xe plasma FIB. The distribution of Fe and Mn is shown where a 6.5 at.% Mn iso-surface is used to show the phase boundary. b) shows the mass spectrum obtained from the measurement of the deuterated specimen. c) shows the top view of the reconstruction where also a C-rich linear feature is shown using a 0.5 at.% C iso-surface. d) shows the one-dimensional concentration profile along the direction highlighted by the black in c).*

## Conclusion

Here, we investigated the possibility of site-specific specimen preparation for analysing H/D in APT, using gaseous charging. We revealed that electron and ion beam irradiation led to the formation of vacancy clusters sub-surface, which acted as a strong trap for H/D, enriched by up to 35 at.% D. These irradiation-induced damages cannot be overlooked especially when investigating H/D by using APT. By sequentially adjusting steps in the specimen preparation workflow, the density of these vacancy clusters was progressively minimized, which required eliminating TKD, and ion beam deposition of Pt protective layer, as well as the Ga$^+$ ion beam. By using a Xe plasma FIB, and the e-beam for the Pt-deposition, the radiation-induced damage was successfully eliminated. This preparation technique allowed us to probe the H/D traps, inherent to the material where we identified a carbon-rich dislocation core, in ferrite, that was enriched with up to 0.3 at.% D.

## Acknowledgement

The authors gratefully acknowledge Uwe Tezins, Andreas Sturm and Christian Bross for their support to the FIB and APT facilities at MPIE. A.S. acknowledges the financial support from Deutsche Forschungsgemeinschaft (DFG) under the collaborative research centre SFB/TR 103 and support from research fund for coal and steels (RFCS) under the project HYDRO-REAL-899335. H.K. and B.G. acknowledge the financial support from ERC-CoG-SHINE-771602.